\documentclass[twocolumn]{aastex63}

\submitjournal{ApJ}

\shorttitle{Cetus Model}
\shortauthors{Chang et al.}
\begin{document}

\title{Is NGC 5824 the core of the progenitor of the Cetus Stream?}

\correspondingauthor{Zhen Yuan}
\email{sala.yuan@gmail.com}

\author{Jiang Chang}
\affiliation{Key Lab of Optical Astronomy, National Astronomical Observatories, CAS, 20A Datun Road, Chaoyang District, 100012 Beijing, China}
\affiliation{Purple Mountain Observatory, CAS, No.10 Yuanhua Road, Qixia District, Nanjing 210034, China}

\author{Zhen Yuan}
\affiliation{Key Laboratory for Research in Galaxies and Cosmology, Shanghai Astronomical Observatory, Chinese Academy of Sciences, 80 Nandan Road, Shanghai 200030, China}

\author{Xiang-Xiang Xue}
\affiliation{Key Lab of Optical Astronomy, National Astronomical Observatories, CAS, 20A Datun Road, Chaoyang District, 100012 Beijing, China}

\author{Iulia T. Simion}
\affiliation{Key Laboratory for Research in Galaxies and Cosmology, Shanghai Astronomical Observatory, Chinese Academy of Sciences, 80 Nandan Road, Shanghai 200030, China}

\author{Xi Kang}
\affiliation{Zhejiang University-Purple Mountain Observatory Joint Research Center for Astronomy, Zhejiang University, Hangzhou 310027, China}
\affiliation{Purple Mountain Observatory, CAS, No.10 Yuanhua Road, Qixia District, Nanjing 210034, China}

\author{Ting S. Li}
\affiliation{Observatories of the Carnegie Institution for Science, 813 Santa Barbara St., Pasadena, CA 91101, USA}
\affiliation{Department of Astrophysical Sciences, Princeton University, Princeton, NJ 08544, USA}
\affiliation{NHFP Einstein Fellow}

\author{Jing-Kun Zhao}
\affiliation{Key Lab of Optical Astronomy, National Astronomical Observatories, CAS, 20A Datun Road, Chaoyang District, 100012 Beijing, China}

\author{Gang Zhao}
\affiliation{Key Lab of Optical Astronomy, National Astronomical Observatories, CAS, 20A Datun Road, Chaoyang District, 100012 Beijing, China}

\begin{abstract}

The complicated story of the Cetus Stream (CS) is recently revealed by its newly discovered $\sim$150 members with 6D kinematics from the cross-matched catalog of LAMOST DR5 K giants and Gaia DR2. It exhibits a very diffuse structure at heliocentric distances between 20 to 50 kpc, extending over at least 100 degrees, and crossing the Galactic plane. Interestingly, The CS is dynamically linked to a massive globular cluster, NGC 5824. A suggestive scenario is that NGC 5824 was the nuclear star cluster of the dwarf progenitor of the CS. We explore this scenario by modeling the disruption process of a dwarf galaxy in the Milky Way potential, on the orbit of NGC 5824, using a suite of N-body simulations. Our results show that the simulated stream can marginally recover the main component of the CS, which is the densest part of the observed stream. Inspired by this mismatch, we use a dwarf progenitor following the representative orbit of the main component members, and find it can reproduce the general morphology of the CS. This gives us a more favorable scenario of the CS progenitor, in which NGC 5824 was not the core, but located off-center. Our fiducial model also predicts a vast extension of the CS in the South, surprisingly coincident with a newly discovered wide Southern stream ``Palca''. Another more diffuse substructure, the Eridanus-Pheonix overdensity is also likely to be related to the CS progenitor.

\end{abstract}


\section{Introduction}
\label{intro}


The standard $\Lambda$CDM cosmological model predicts that the Milky Way (MW) formed and grew through galactic mergers. In this hierarchical growth picture, the inner stellar halo consists of early accreted systems that have sunk into the center of the galaxy \citep[see e.g.][]{bullock05,cooper10,amorisco17}. Their stellar components have gone through severe phase mixing, and constitute the stellar halo with a relatively smooth distribution in configuration space. Thanks to the precise proper motion measurements from Gaia DR2 \citep{gaia18, gaiadr2}, a great deal of debris in the nearby halo are discovered from both kinematical and dynamical spaces, including those from two massive accreted dwarf satellites, Gaia-Enceladus-Sausage and Sequoia \citep[see e.g.][]{belokurov18,helmi18, myeong19, matsuno19}, and those potentially from minor mergers \citep[][]{myeong18c, kopp19, yuan20}, which need chemical abundances measurements to confirm their unique origins.

As we move further from the Galactic centre, at distances larger than 10 kpc, the un-mixed portions of these large early events start to appear in the shape of stellar clouds splattered across huge regions of the sky e.g. Herqules-Aquila and Virgo Clouds \citep{simion19}. The outer halo is also the reservoir of the wreckage from late minor mergers. The dynamical friction is not strong enough for low mass dwarf satellites to lose orbital energy efficiently, leaving them confined in the outer region of the Galaxy \citep[see e.g.][]{amorisco17}. Stars are continuously being stripped from these small systems as they orbit around the host galaxy, which form streams with relatively coherent distributions in phase space shown as overdensities; the most famous example of stream in the Milky Way is the Sagittarius (Sgr) Stream \citep[see e.g.,][]{mateo96,ibata01,majewski03,belokurov06,ibata20,antoja20,ramos20}.

Stellar streams inherit, in their rich observational features, the memory of their progenitors and host galaxy environment, rendering them valuable fossils in our study of ancient dwarf galaxies, assembly history of the MW and present day Galactic potential \citep[see e.g.][]{pw16,bonaca19a,bonaca19b,malhan20}. For instance, the Sgr stream is a textbook example of an interaction between a host galaxy and a satellite dwarf galaxy. Several studies have attempted to model the Sgr dwarf and the its impact on the disc with N-body simulations. Although the full picture of the merging event is far from clear, some features of the Sgr tidal stream can be successfully explained, such as the bifurcations in the leading tail \citep{penarrubia10}, the confliction between angular position and radial velocity \citep{law10}, and its footprint in phase-space \citep{dierickx17}. \citet{laporte18, laporte19b} showed that the Sgr dwarf can induce the Disc phase-space spiral \citep{antoja18} seen by Gaia DR2 \citep{gaiadr2} by exciting the disc during its first pericentric passage, and it can also reproduce some of the low latitude overdensities such as Monoceros Ring and Anticenter Stream \citep{laporte19a,laporte20}. \citet{vasiliev20} analyzed the stellar kinematics in the Sgr core using a large suite of N-body simulations tailored to the observational constaints. 

However, the Sgr stream is the only stream stripped from a dwarf galaxy that has coherent structure in configuration space, and confirmed MW globular clusters (GCs) associations until our recent work on the Cetus Stream (CS) \citep{yuan19}. We determined that the CS is dynamically associated with NGC 5824. and also updated the member list of the Cetus Stream from $\sim$ 40 candidates \citep{yam13} to 151 confirmed ones, based on their orbital properties. About half of the members are located in newly discovered components of the CS which was not identified in the original discovery paper \citep{newberg09}. The current CS members with the full 6D kinematics have the average orbital inclination angle of $\sim$ 60$^{\circ}$, significantly different from 87$^{\circ}$ in \citet{yam13}. The latter was obtained from the fittings to the gradients of line-of-sight velocity and distance in Galactic latitude $b$, from $\sim$ 40 candidate members, based on which the numerical modeling of the CS was built. Therefore, given the updated members with precise orbital properties for the first time, it is very necessary to rebuild the formation process of the CS, and this work is the first attempt in this regard.

One of the important unanswered questions about the CS dwarf progenitor is whether NGC 5824 is its core. This scenario was first postulated based on line-of-sight velocity information only from the discovery paper \citep{newberg09}, and followed by a number of studies which investigated the spread in Magnesium abundances of NGC 5824 \citep{roederer16, mucciarelli18}, and the extended structures around the GC with deep photometry \citep{walker17,kuzma18}. In this work we seek to answer this question with dedicated N-body modeling of the merging event. In Sec.~\ref{sec:obs}, we summarize the key observational features of the CS, which are critical for comparisons with simulations. The numerical framework is set up in Sec.~\ref{sec:sim}. We assume two different scenarios for the progenitor, described in Sec.~\ref{subsec:s1} and \ref{subsec:s2}. Guided by the distinctive features of the observed CS in 6D phase space, our goal is to verify if NGC 5824 is the core of its disrupted progenitor. Note that quantitative fittings to the observed CS is still difficult given its diffuse nature and the current number of members with kinematic information, thus our aim is to obtain a favorable scenario which can match with the general morphology of the CS. Based on this, we predict the Southern extension of the CS, and surprisingly find it is coincident with the newly discovered Palca Stream \ref{subsec:south}. The relationship between the Cetus progenitor and NGC 5824 is discussed in Sec.~\ref{sec:diss}, as well as the effects on the stream from varying initial conditions and the environment of the disruption process. The final conclusions are given in Sec.~\ref{sec:con}.

\begin{figure*}
\centering
\includegraphics[width=\linewidth]{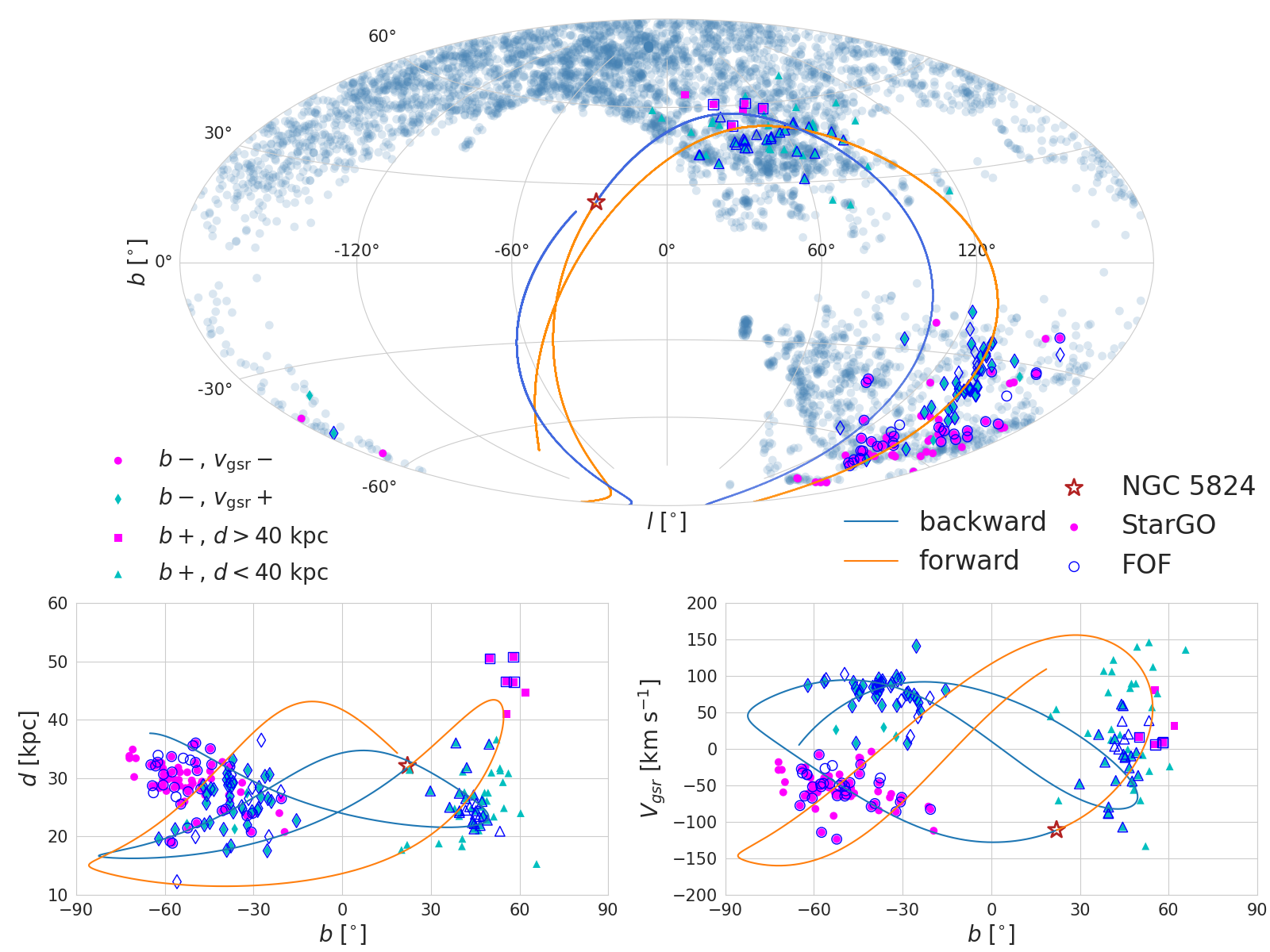}
\caption{The overview of the observational constraints of the CS. The CS members identified by \textsc{StarGO} from \citet{yuan19} are plotted by solid symbols, and those found by \textsc{FoF} are plotted by open blue symbols. Upper: The on-sky projection of the CS, superposed on the combined sample of LAMOST K Giants and SDSS BHBs, shown as steel-blue circles. Bottom: The CS in the ($b$, $d$) and ($b$, $v_{\rm gsr}$) spaces, which are decomposed into four components , denoted by different colors and symbols. The more densely populated Southern part of the CS, compared to the Northern part, is divided into the ``main'' component (magenta circles), and the ``+ $v_{\rm gsr}$'' component (cyan diamonds). The red star symbol denotes NGC 5824, whose orbit (forward: orange; backward: steel-blue) goes through the CS members in all three spaces.}
\label{fig:obs}
\end{figure*} 

\section{The Observed Cetus Stream}
\label{sec:obs}
We present the overview of the observed CS in Galactic coordinates as shown in Fig.~\ref{fig:obs}. In the upper panel, the blue background exhibits the footprint of the combined sample of LAMOST K Giants \citep{liu14} and SDSS BHBs \citep{xue08, xue11} used for stream searching in \citet{yuan19}. This sample clearly has the emptiness around the disc region ($|b|<15^{\circ}$) due to the limited sky coverage of both surveys. The CS members identified by \textsc{StarGO} from \citet{yuan19} are plotted by solid symbols, and the members obtained by a \textsc{Friends-of-Friends}-based method from Xue et al. (in prep) are plotted by open blue symbols. We can immediately see that the CS has very diffuse structure, and two parts on sky projection, separated by the Galactic disc. The original discovery of the CS is located in the South \citep{newberg09}, and its northern counterpart is recently found by \citet{yuan19}, which is less densely populated. As shown in \citet{yuan19}, the integrated orbit of NGC 5824 (dark red star symbol) transverse both parts of the CS. The steel-blue and orange lines denote its forward and backward orbits respectively.

The orbit of NGC 5824 also passes through the CS members in ($b$, $d$) and ($b$, $v_{\rm gsr}$), the two signature spaces where the CS was originally identified, shown in the bottom two panels of Fig.~\ref{fig:obs}. $v_{\rm gsr}$ is the line-of-sight velocity transformed in Galactic Standard of Rest (GSR) frame, which moves at the velocity of the solar with respect to the Galactic center. The typical velocity error is about 5 km s$^{-1}$, and the distance uncertainty is 13$\%$ for K Giants and 5$\%$ for BHBs \citep[see][and reference within]{yuan19}. From the CS members identified by \citet{yuan19}, the Southern CS is clearly separated into two clumps in ($b$, $v_{\rm gsr}$), shown in the right panel. The members in the region of the original CS \citep{newberg09} (magenta circles) have negative $v_{\rm gsr}$, and negative gradient in ($b$, $d$), which is the major population identified by \citet{yuan19}, referred to as the ``main'' CS component. Those stars with positive $v_{\rm gsr}$ is a new component (cyan diamonds) and referred to as ``+ $v_{\rm gsr}$'' component, generally following a positive gradient in ($b$, $d$). These are the two relatively densely populated parts of the CS, providing essential constraints on the modelings of the stream, discussed in detail in Sec.~\ref{sec:res}. Compared with these Southern components, the Northern part is much more diffuse, denoted by magenta squares ($d>$ 40 kpc) and cyan triangles ($d<$ 40 kpc).

NGC 5824 was first postulated to be associated with the CS by \citet{newberg09} based on line-of-sight velocity information. This association is confirmed by \citet{yuan19} according to their very similar orbital properties from the newly available 6D phase space information. The orbit of NGC 5824 is in line with the distribution of the CS members in different spaces of Fig.~\ref{fig:obs}, leading to an obvious possibility that the GC was the core of the progenitor of the stream.

\begin{figure*}
\centering
\includegraphics[width=\linewidth]{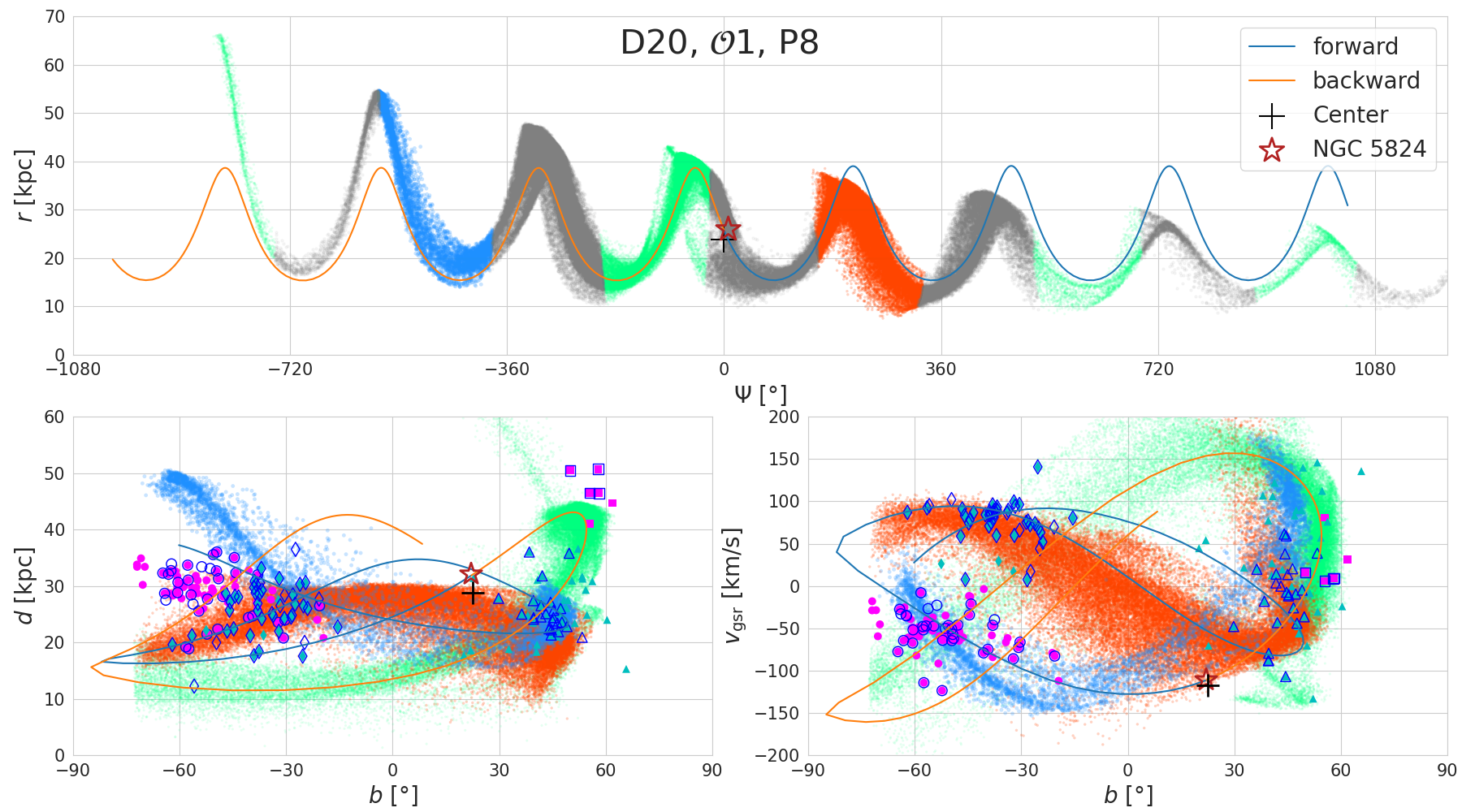}
\caption{The simulated stream stripped from the dwarf progenitor on the orbit of NGC 5824 in Scenario I. Upper: The entire simulated stream from the run ($\mathcal{O}1$, D20, P8) is unfolded along the azimuthal angle $\Psi$. The stream is shown as the gray scatter plot, where the wraps entering the observable sky are colored in light green. The two wraps closest to the Southern CS are further highlighted in blue and orange. NGC 5824 (red star) overlaps with the center of the disrupted dwarf progenitor (black cross). The integrated orbit of NGC 5824 is plotted for the guidance. Bottom: Comparisons between the observed CS and the simulated stream in the ($b$, $d$) and ($b$, $v_{\rm gsr}$) spaces. The ``+ $v_{\rm gsr}$'' component can be recovered by the orange wrap, but the ``main'' component is in disagreement with the blue wrap: It has a steeper slope in ($b$, $d$), and is also too thin to be consistent with the ``main'' component.}
\label{fig:o1}
\end{figure*}

\section{Simulation Setup}
\label{sec:sim}

We set up the numerical framework of modelling the disruption of the Cetus progenitor in this section. The MW model used is \texttt{MWPotential2014} from \citep{bovy15}. It includes both dark matter and stellar components. Specifically, the dark matter halo of the MW is described as a spherically symmetric Navarro-Frenk-White (NFW, \citet{navarro97} potential with virial mass $M_{\rm vir}= 10^{12}$M$_{\odot}$, consistent with the recent studies of the MW mass \citep[see e.g.][and references within]{li19, wang19}. The stellar component of the MW contains a Miyamoto-Nagai disk \citep{miyamoto75} of $6.8\times10^{10}$M$_{\odot}$ with scale length $r_{\rm d}= 2.4$ kpc, and a Hernquist bulge of $0.5\times10^{10}$M$_{\odot}$ with scale length $r_{\rm b}=0.2$ kpc \citep{bovy13}. The dwarf progenitors are modeled as particles with varying initial conditions. 

The Cetus progenitor is estimated to have a stellar mass of 10$^4$M$_{\odot}$ -- 10$^6$M$_{\odot}$ \citep{yuan19}, based on the mass metallicity relationship of surviving dwarf satellite galaxies of the MW \citep{kirby13}. However, this relationship has large uncertainties in the low metallicity region, and more importantly, it may not be applied to the disrupted dwarf galaxy population of the MW. Therefore, we set the stellar mass of the progenitor as a free parameter, which scales with the total mass at its apo-center passage, when the outer part of the dark matter halo has been substantially stripped \citep{chang13, penarrubia08}. Specifically, the stellar to halo mass ratio is set to be 0.1, as a conventional value obtained from the similar modelings of the Sgr stream \citep{penarrubia10, dierickx17}. Given the fact that the CS is much less massive than the Sgr Stream, we set the CS dwarf progenitor at least one order of magnitude smaller than the Sgr progenitor, with a typical total mass of $\lesssim$10$^9$M$_{\odot}$. It is less than a few percent of the MW mass, therefore the dynamical friction has negligible effect on the progenitor's orbit \citep[see e.g.,][]{binney08, mo10,vdb18}. We describe the progenitor dwarf with dark matter and star particles using the modified \textsc{Gadget3} \citep{springel05}, and let it evolve in the analytic MW potential for several orbital periods. We explore a variety of initial conditions in Sec.~\ref{subsec:ini}. In each run, we use a similar recipe for the dwarf satellites according to \citet{chang13}. For simplicity, the concentration parameter for all dwarf satellites is set to be 15 \citep[see e.g.][]{springel08, dutton14}. The fiducial satellite dwarf has a total mass of 2$\times$10$^9M_{\odot}$, and a relatively large stellar disc with a half light radius of 1.6 kpc.

\section{Result}
\label{sec:res}
\subsection{Scenario I: NGC 5824 is the core}
\label{subsec:s1}

We first explore the scenario that NGC 5824 is the core of the CS dwarf progenitor, in which case, these two share the same orbit, referred to as $\mathcal{O}$1. After the system evolves a sufficient long time of about eight apocentric passages ($\sim$ P8), we show the resulting stream by unfolding it along the azimuthal angle $\Psi$, shown as the gray scatter in the upper panel of Fig.~\ref{fig:o1}. $\Psi$ is the angle between the star and the progenitor's center, with respect to the MW center \citep[see e.g.][Chapter 3.1]{binney08}. It is used to indicate a star's location on the stream, which records its orbital phase along $\mathcal{O}$1. The wraps entering the observable sky with the avoidance of the disk region are colored in light green, which are not covered by the halo sample used, i.e. Dec.$>-$10$^{\circ}$, $|b|>15^{\circ}$. The two wraps critical to compare with the observations, are highlighted in blue and orange.

All the wraps entering the observable sky are plotted in the ($b$, $d$) and ($b$, $v_{\rm gsr}$) spaces shown in the bottom two panels of Fig.~\ref{fig:o1}. As we can see, the blue wrap, as the most promising wrap to recover the ``main'' component, is too thin to give a reasonable match. Moreover, its gradient in ($b$, $d$) is far more steeper than that of the observed ``main'' component. The  ``+ $v_{\rm gsr}$'' component can be reproduced by the orange wrap, and the Northern members are generally coincident with the mixture of all the colored wraps. However, the disagreement on the ``main'' component disfavors this scenario of NGC 5824 being the core. We find that this mismatch cannot be reconciled by a longer stream from a more massive progenitor dwarf or longer evolution time, implying that it results from the inconsistent orbital phase between NGC 5824 and the center of the CS.

\begin{figure}
\centering
\includegraphics[width=\linewidth]{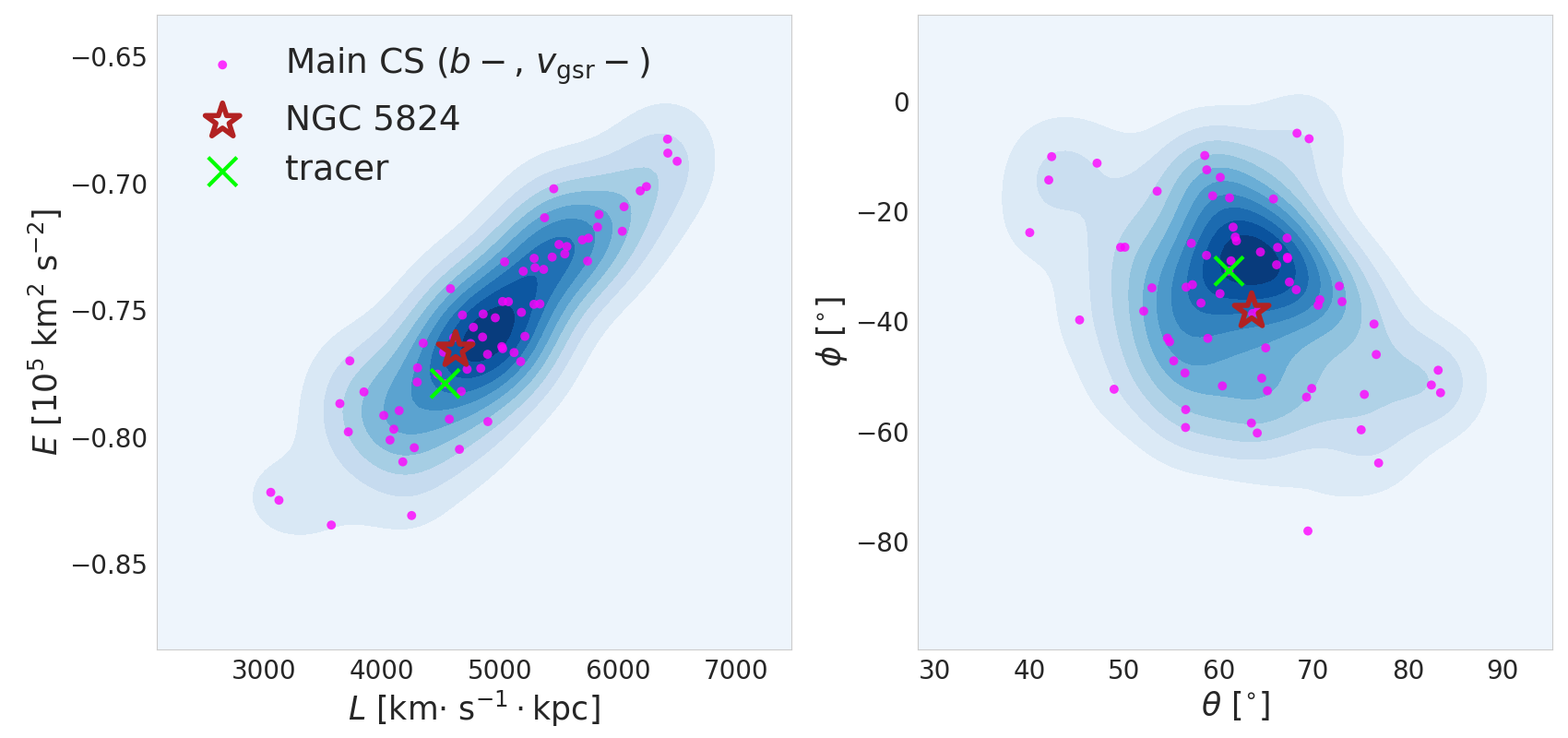}
\caption{The ``main'' CS members in the ($E$, $L$) space (left) and the ($\theta$, $\phi$) space (right), shown as both scatter and contour plots. The latter space represents the directions of angular momentum. The tracer star is plotted by the lime-green ``X'', which is located in the low-$E$ region and almost at the center of the density peak in ($\theta$, $\phi$). }
\label{fig:le}
\end{figure}

\begin{figure*}
\centering
\includegraphics[width=\linewidth]{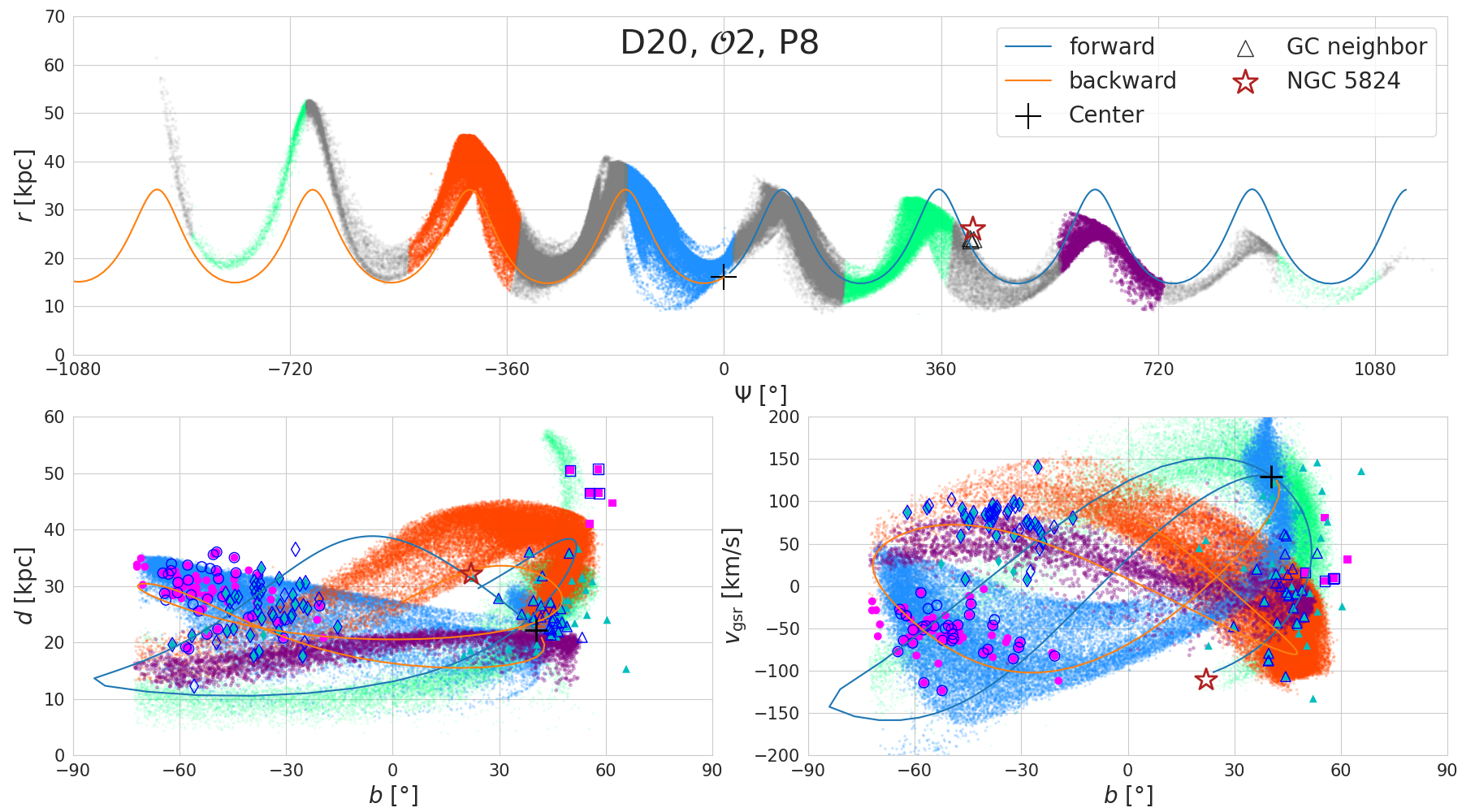}
\caption{The simulated stream stripped from the dwarf progenitor on the typical orbit from ``main'' component in Scenario II, similar to Fig.~\ref{fig:o1}. Upper: The entire simulated stream from the run ($\mathcal{O}2$, D20, P8), and the observable portion is colored in light green . The three wraps closest to the Southern CS are further highlighted in blue, orange, and purple. The integrated orbit of the tracer is plotted for the guidance. Bottom: Comparisons between the observed CS and the simulated stream in the ($b$, $d$) and ($b$, $v_{\rm gsr}$) spaces. Both the purple and orange wraps contribute to he ``+ $v_{\rm gsr}$'' component, and the blue wrap can well reproduce the ``main'' component. }
\label{fig:o2}
\end{figure*}

\subsection{Scenario II: A Typical Main Stream Member as the Tracer}
\label{subsec:s2}

Since the Cetus progenitor on the orbit of NGC 5824 cannot well reproduce the observed CS, we explore other possible tracers for this stellar-debris system. As we have shown in Sec.~\ref{sec:obs}, the ``main'' component is the most densely populated region of the stream. It can naturally lead to an assumption that the stream members in the ``main'' component are close to the center of the dwarf progenitor before its disruption. Based on this, we first pick out the ``main'' component, by selecting stream members with $b<0^{\circ}$ and $v_{\rm gsr}<0$ km s$^{-1}$, yielding 68 stars. Their distribution in angular momentum ($L$) versus orbital energy ($E$) space is shown as the scatter and contour plots, in the left panel of Fig.~\ref{fig:le}. The direction of angular momentum vector is characterized by two angles $\theta$ and $\phi$, which are defined as $\theta=\rm{arccos}$($L_{\rm z}$/$L$) and $\phi=\rm{arctan}$($L_{\rm x}$/$L_{\rm y}$), where $L_{\rm x}$, $L_{\rm y}$and $L_{\rm z}$ are the $x$, $y$ and $z$ components of angular momentum vector in Galactic coordinate. The distribution of the ``main'' members in the ($\theta$, $\phi$) space is shown in the right panel, which has a density peak around (65$^{\circ}$, $-$30$^{\circ}$) as seen from the contour plot. Based on our assumption, the center of dwarf progenitor would have ($\theta$, $\phi$) values close to the density peak. We select the stars in the peak vicinity and integrate their orbits for about three orbital periods both forward and backward, and find that they all follow very similar trajectories, which can traverse most of the CS members in 6D phase space. We therefore choose a representative star as the tracer star and name its orbit as $\mathcal{O}$2. The tracer star marked by the green cross in Fig.~\ref{fig:le}, is located in the low-$E$ region of the entire distribution of the ``main'' members, consistent with the assumption that the tracer is close to the center of the dwarf progenitor. There is a noticeable distance between NGC 5824 and the tracer star in ($\theta$, $\phi$), implying that these two tracers would result in different progenitor's orbits, and hence different streams.

The entire stream on $\mathcal{O}$2 with eight apocentric passages, is plotted in the upper panel Fig.~\ref{fig:o2} in the same way as in Fig.~\ref{fig:o1}. The wraps that enter the observable sky are colored in light green, and three of them are highlighted in blue, orange and purple, which are used to match with the two critical components of the observed CS. In contrast to Scenario I, the blue wrap is densely populated in region of the ``main'' component. Both the orange and the purple wraps can contribute to the ``+ $v_{\rm gsr}$'' component. The Northern members are covered by the mixture of all the wraps similar to that in Scenario I. Note that we shift the last snapshot of the simulation by 0.2 Gyr , which is equivalent to adjusting the orbital phase of the progenitor's center, in order to have better matching with the observations. Overall, the stream on $\mathcal{O}$2 matches the observed CS much better. Therefore, the typical ``main'' stream member as the tracer, gives a more favorable scenario of the progenitor system. 

\begin{figure*}
\centering
\includegraphics[width=\linewidth]{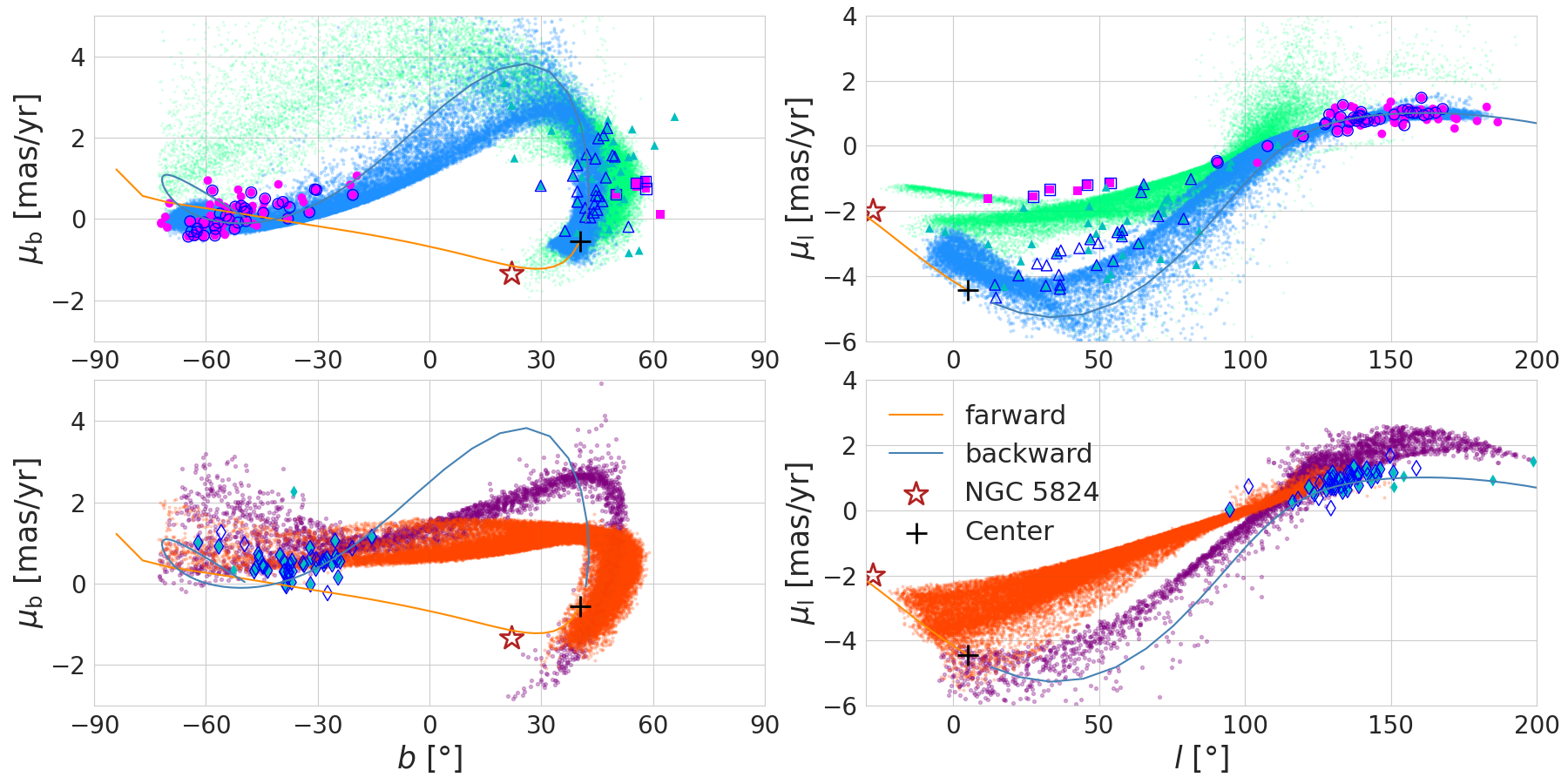}
\caption{Comparisons between the observed CS and the simulated stream of Scenario II in the ($b$, $\mu_b$) space (left columns) and the ($l$, $\mu_l$) space (right columns), similar to Fig.~\ref{fig:o2}. Upper:  The blue wrap match well with the ``main'' component (magenta circles), located at $b$ = ($-$30$^{\circ}$, $-$60$^{\circ}$), $l$ = (100$^{\circ}$, 200$^{\circ}$). Bottom: The orange and purple wraps are in good match with the  ``+ $v_{\rm gsr}$'' component (cyan diamonds). The Northern members are in agreement with multiple wraps. }
\label{fig:pm}
\end{figure*}

We then compare the simulated stream in this scenario with the observed CS in the ($b$, $\mu_b$) and ($l$, $\mu_l$) spaces shown as Fig.~\ref{fig:pm}, using the recently available proper motion measurements from \citet{gaiadr2}, with typical error of 0.1 mas yr$^{-1}$ for K giants and 0.3 mas yr$^{-1}$ for BHBs. The blue wrap in the upper panels, agrees very well with the ``main'' component (magenta circles) located at $b$ = ($-$30$^{\circ}$, $-$60$^{\circ}$), $l$ = (100$^{\circ}$, 200$^{\circ}$). We plot the  ``+ $v_{\rm gsr}$'' component (cyan diamonds) in the lower panels, which are in good match with the orange and purple wraps. Consistently, the Northern members can be recovered by multiple wraps. We refer to this run as the fiducial run, and discuss its implications in Sec.~\ref{subsec:south} and Sec.~\ref{subsec:gc}. A suite of runs with varying initial conditions on the basis of the fiducial settings are discussed in Sec.~\ref{subsec:ini}.

\subsection{Cetus in the South}
\label{subsec:south}

\begin{figure*}
\centering
\includegraphics[width=\linewidth]{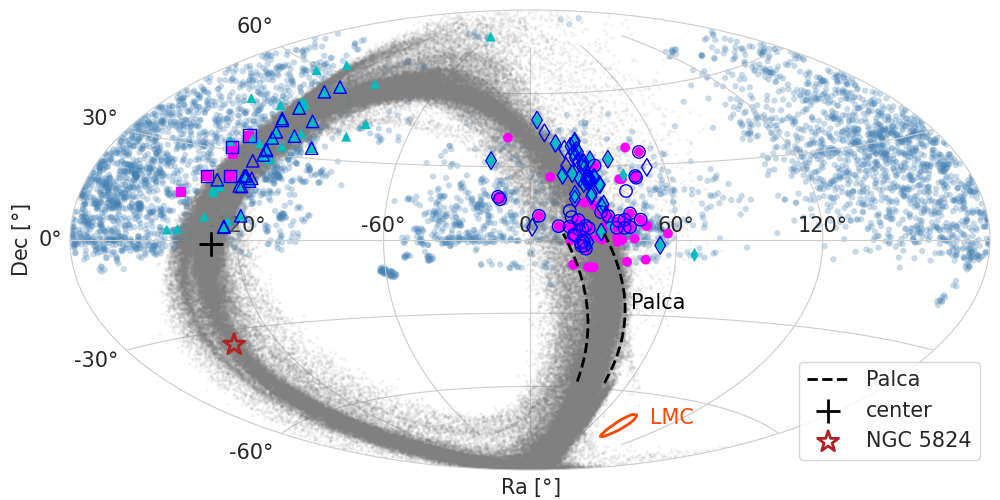}
\caption{The Cetus Stream in Equatorial coordinates shown in the same way as in Fig.~\ref{fig:obs}. The entire simulated stream is plotted by gray scatter, and the center of the disrupted progenitor is shown as black ``+''. The black dashed lines denote the Palca stream discovered by \citet{shipp18}, which matches well with part of the simulated stream in the Southern Sky.}
\label{fig:radec}
\end{figure*}

\begin{figure}
\centering
\includegraphics[width=\linewidth]{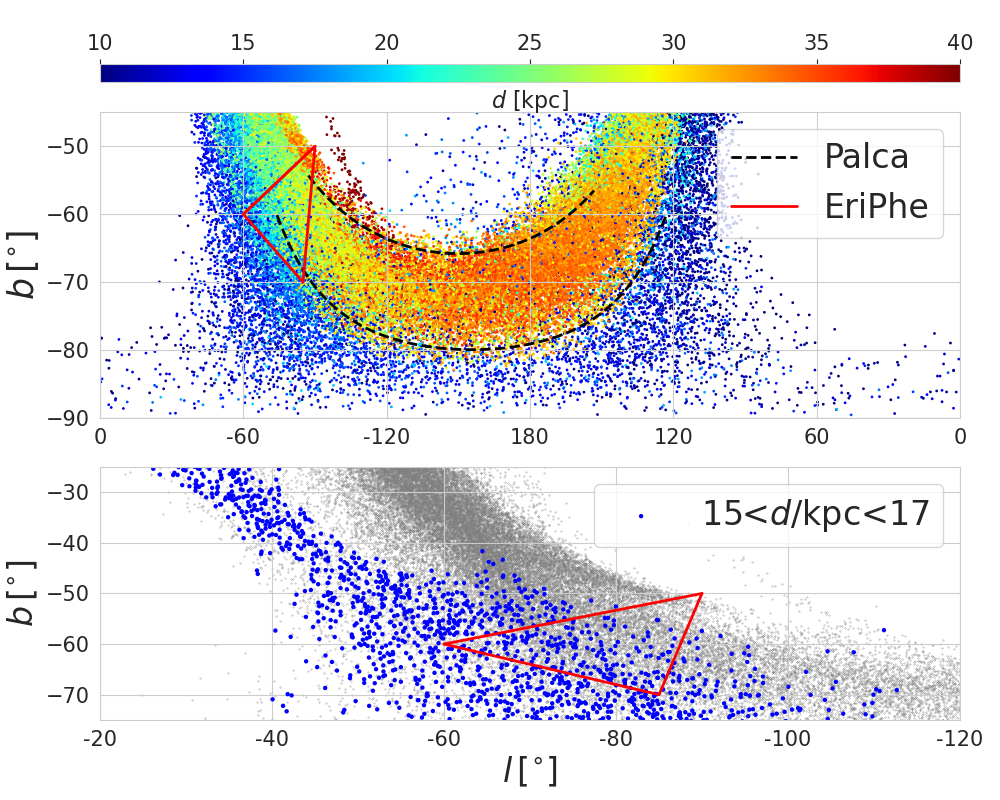}
\caption{The footprint of the fiducial stream in the selected sky area in the Galactic South. Top: The section of the stream with $b\leqslant-$45$^{\circ}$, where the color bar denotes the heliocentric distance. The black dashed line represents the footprint of the Palca Stream at $d\approx$ 36.3 kpc, which matches well with the trajectory of the fiducial stream as well as its distance distribution. The red triangle denotes the region of Eri-Phe overdensity with $d\sim15$ -- 17 kpc. Bottom: The zoom-in region of the stream near Eri-Phe marked by the same red triangle as the top panel, where the star particles in the distance range (15 kpc -- 17 kpc) are highlighted in blue.}
\label{fig:south}
\end{figure}

The on-sky projection of the fiducial stream is shown in heliocentric coordinates in Fig.~\ref{fig:radec}. Clearly, about half of the stream is left in the Southern Sky. We are curious to see if any footprint of the CS can be traced there, which is not covered by the spectroscopic data used in this study. Thanks to the deep photometric data acquired by the Dark Energy Survey \citep[DES;][]{des16}, numerous streams were discovered \citep{shipp18}. We find that a wide stream, ``Palca'', depicted by the black dashed lines in Fig.~\ref{fig:radec}, perfectly matches the predicted stream in the south of the ``main'' CS. We then zoom into the sky region of the Palca stream ($b<-$45$^{\circ}$), shown in the upper panel of Fig.~\ref{fig:south}, where the simulated stream is color coded by its distance. Both the length and trajectory agree well with the Palca Stream. According to the results using the matched-filter technique with a grid of distance moduli from \citet{shipp18}, the distance of the Palca is found to be 36.3 kpc. This is consistent with the majority of the star particles with orange shades between the two black dashed lines. Based on these, we conclude that the Palca Stream is very likely the southern extension of the CS. line-of-sight velocities of the stream members are needed to confirm the connection, and they could be soon provided by the Southern Stellar Stream Spectroscopic Survey \citep[S$^5$;][]{s5}. According to the simulations, the CS would extend further to the Southern sky, and reach the sky region of the LMC shown in the lower panel of Fig.~\ref{fig:radec}. This region is beyond the coverage of the DES, but will be covered by LSST \citep{lsst09}. We look forward to verifying these predictions using future photometric and spectroscopic surveys.

Besides the Palca Stream, there is another over density, Eridanus-Phoenix (EriPhe), located close to the simulated stream in the Galactic South, shown as a red triangle in Fig.~\ref{fig:south}. It was discovered earlier by \citet{li16} from the DES. From simulations, there are several wraps across the triangle region with a wide range of distances from 15 kpc (blue) to 25 kpc (light-green). We further highlight the star particles with $d\sim15$ -- 17 kpc in blue in a zoom-in patch of the sky, as shown in the lower panel. The distribution of these blue star particles agrees with the triangle region of EriPhe, whereas the confirmation of their association requires kinematic information. Note that the part of the sky north to the triangle is not in the footprint of the DES according to the original discovery paper \citep{li16}. We need deep photometric observations in the $l\sim280^{\circ}$ -- 340$^{\circ}$, and $b\sim-60^{\circ}$ -- $-30^{\circ}$ sky region, which will become available in the upcoming LSST era. 

\section{Discussions}
\label{sec:diss}

\subsection{The Association with NGC 5824}
\label{subsec:gc}

Scenario II gives the fiducial stream in agreement with the general morphology of the observed CS in 6D phase space. Based on this, we explore the possible relationship between the Cetus progenitor and NGC 5824. To address this problem quantitatively, more dedicated modeling of a globular cluster in a disrupted dwarf galaxy is required, which is beyond the scope of this paper. The aim of this work is to provide some clues about the location of NGC 5824 in the CS progenitor before its infall for future detailed studies.

In the fiducial scenario, we pin down the azimuthal angle of the GC ($\Psi=404^{\circ}$) by finding its closest point on $\mathcal{O}$2 in 6D phase space. It is plotted as the dark red star symbol in the upper panel of Fig.~\ref{fig:o2}, with the current Galactocentric radius of 27 kpc. For the consistency check, we select the star particles from the fiducial run that are close to the current GC satisfying $\Delta v\lesssim20$ km s$^{-1}$, and $\Delta r \lesssim5$ kpc. The three closest neighbor particles are plotted as black triangles, which consistently have almost the same $\Psi$ as that inferred for NGC 5824 from the integrated orbit. The GC is about 1.5 wraps ahead of the center of the progenitor, in line with the assumption that the GC is situated off-center during disruption.

This immediately leads to another question that where was NGC 5824 located in the dwarf progenitor just before its infall. We can give suggestive solutions by tracing the three neighboring stars closest to the GC, which have a large range of distances (1 -- 5 kpc) with respect to the progenitor's center. Given that NGC 5824 is very massive (10$^6$M$_{\odot}$), the dynamical friction timescale in a dwarf galaxy with a mass of 2$\times$10$^9$M$_{\odot}$ (D20) is around 1 Gyr as estimated by \citet{yuan19}. Since the GC is very old, with the age of 12 Gyr, it would have been dragged down to the center before the system fell into the MW. As predicted from the simulations, when the CS progenitor first crossed the apo-center around 5 Gyr ago, the GC was not the nuclear star cluster. How to prevent such a massive cluster from sinking into the center in such a long timescale ($\sim$7 Gyr) remains a puzzle. A similar timing problem exists in the Fornax dSph, where three of five GCs are located outside its half-light radius, and would sink to the center within a few Gyr after the birth. There are several plausible solutions to this problem. One is the orbital expansion caused by the strong tidal effect from the MW \citep{oh00}. The updated Fornax proper motion shows that it never comes close enough to the MW, which rules out this possibility. However, this is still a possible solution for the Cetus progenitor, which suffers stronger tidal disruption compared to the Fornax dSph. Even though we simulate its evolution in the MW potential for eight passages, the tidal effect before the first apo-center crossing is difficult to estimate, but it could cause an orbit expansion and delay the sinking of the GC. Another promising solution is the `core-stalling' effect caused by the dark matter core according to several studies \citep[e.g.,][]{goerdt06,read06,cole12}. The current positions of Fornax GCs can be explained by a suite of simulations, which show that GCs will remain at the edge of the core radius of the dark matter halo of the Fornax dSph for a sufficiently long time. Although we conclude that the GC is off-center in the CS dwarf progenitor, to study its pre-infall location, we will take into account its dynamical evolution in the progenitor in the followup studies.

An alternative scenario of this stellar-debris system, that naturally avoids the timing problem remains possible, in which case, NGC 5824 belongs to a different dwarf progenitor with very similar orbit as that of the CS progenitor. From a simple trial run, the dwarf progenitor of the GC would leave a stream around the current NGC 5824 with a heliocentric distance range between $\sim$20 -- 40 kpc in the Southern Sky. The part of the stream with distances larger than 30 kpc is prominent, and cannot be made by the Cetus progenitor. To discriminate between the fiducial and this alternative scenario, the verification of the existence of the stream members beyond 30 kpc around NGC 5824 becomes critical. This study will become possible with the LSST \citep{lsst09}, which will cover the region of the sky around NGC 5824 with deep photometric data.

\subsection{Initial Conditions of the Cetus Progenitor}
\label{subsec:ini}

In the fiducial run, the CS dwarf is set to have a mass of 2$\times$10$^9M_{\odot}$ and a diffuse disc with scale length of $r_{1/2}$ = 1.6 kpc. We also explored the effects on the simulated streams from different initial conditions of the dwarf progenitor. This is done by varying parameters from the fiducial settings by small amounts, without trying to explore the entire parameter space. Specifically, we tested two different masses of the dwarf progenitor (20: 2$\times$10$^9M_{\odot}$; 4: 4$\times$10$^8M_{\odot}$) and two types of morphology (D/d: disc; S: spheroidal) as shown in Table~\ref{tab:ic}. The stellar disc in a dwarf progenitor has an exponential density profile characterized by $r_{\rm d}$ and $h_{\rm d}$. D20 and D4 have the same values of specific angular momentum and, d20 has a larger value corresponding to a more compact disc. The spheroidal dwarf progenitor has a Hernquist density profile with scale length $r_{\rm b}$. 

We also compare simulation runs with different evolution time scales, quantified by the number of apo-center passages of the dwarf progenitor. The initial conditions of its kinematics ($R_{\rm init}$ and $V_{\rm init}$) are parameters at its first crossing of four (P4), six (P6), and eight (P8) apo-center passages derived from orbital integrations, shown in Table~\ref{tab:orb}. The last column $T_{\rm inf}$ shows the exact time of a satellite in the MW potential since its first apo-center passage.

\begin{figure*}
\centering
\includegraphics[width=0.9\linewidth]{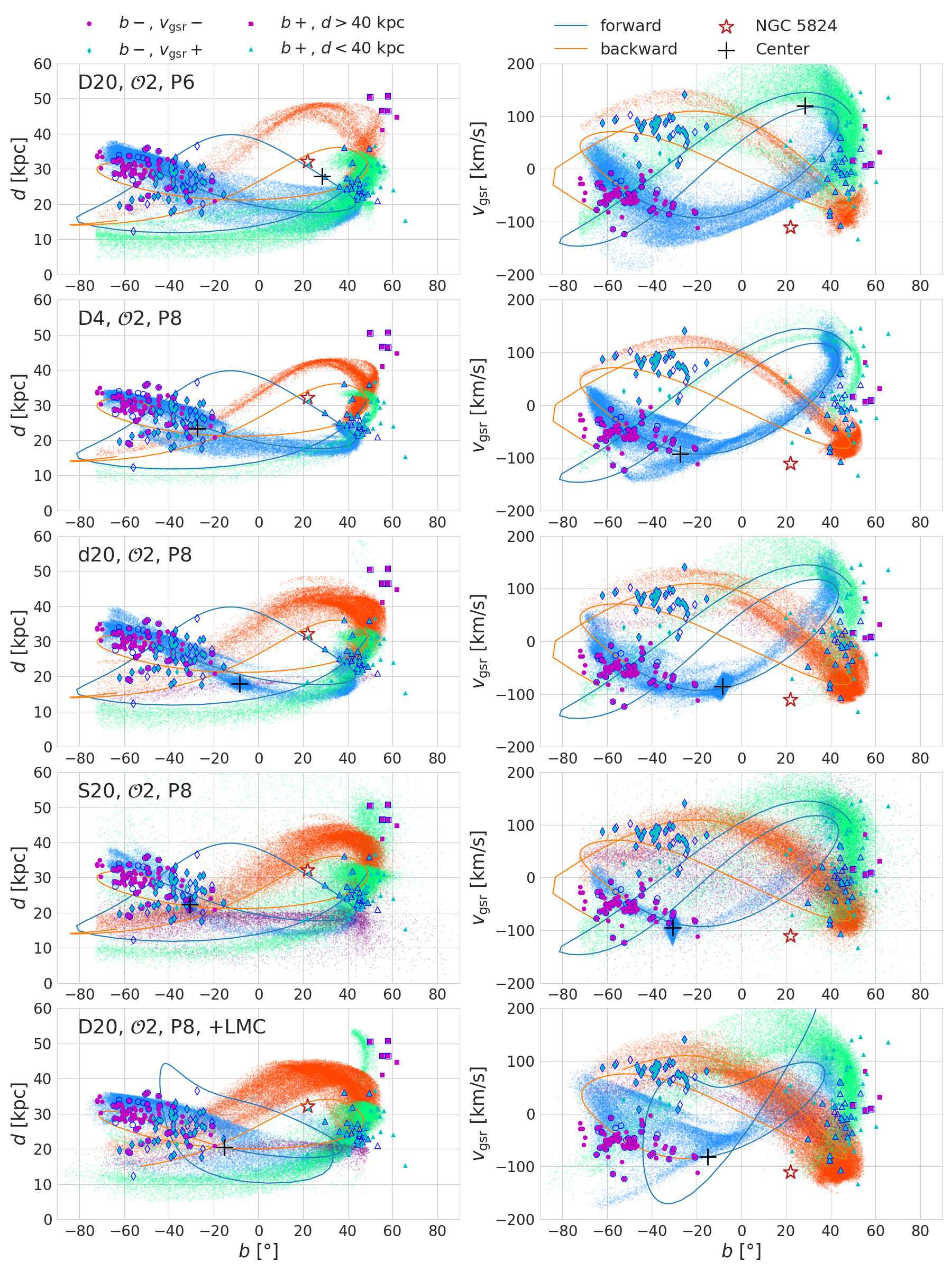}
\caption{The simulated streams from different runs labeled in the legends in the upper left corners. Different wraps entering the observable sky are plotted with distinct colors. The streams from the runs with shorter evolution time (first row) and smaller dwarf satellite (second row) have clear velocity offsets in the ``'+ $v_{\rm gsr}$'' component (cyan diamonds). Both the runs with a more compact disc (third) and spheroidal (forth) have remnant cores around the center of the progenitor. The last run including an LMC with 1.3$\times$10$^{11}$M$_{\odot}$ produces a stream very similar to the fiducial stream, although the forward integrated orbit (blue line) is drastically different from the one without LMC.}
\label{fig:run}
\end{figure*}

We first show the two controlled runs with either a shorter evolution time (P6), or a smaller progenitor mass (D4) in the first two rows of Fig.~\ref{fig:run}. Most of the stream members can be recovered, whereas the matching to the `` + $v_{\rm gsr}$'' component in each scenario has a velocity offset. Compared to the fiducial run, the wraps corresponding to this component from theses two runs are stripped earlier, with larger energy offsets to the progenitor's center, and hence positive velocity offsets to the observations. The fiducial stream is long enough to cover the positive $v_{\rm gsr}$ component with materials stripped later, which does not have such offset in velocity. Therefore, the disagreements of these two runs both come from the fact that these streams are shorter than the fiducial one due to their settings. It is because that the length of a stream mainly depends on the evolution time and the initial spreads in orbital energy of the stellar component, the latter of which scales with the mass of the progenitor \citep[see e.g.][]{johnston01,amorisco15}. We then show the runs of a more compact disc configuration (d20) and a spheroidal profile (S20) in the third and forth rows of Fig.~\ref{fig:run}. The main problem of these two scenarios is the existence of a remnant core due to their relatively compact stellar component, which has not been fully disrupted until the end of the evolution. Since no observational hints of the over-density exist around the current location of the progenitor's center, these two scenarios are not favoured.

From all the simulation runs explored in this work, the most favorable scenario of the Cetus progenitor has a dark matter halo mass of 2$\times10^9$ M$_{\odot}$ at its first apo-center passage, and a long evolution time of eight apo-center passages ($\sim$ 5 Gyr). The stellar component has mass of 2$\times10^8$ M$_{\odot}$, and requires a diffuse disc with $r_{1/2}\sim1.6$ kpc and $M_{1/2}\sim 3.7\times10^8$M$_{\odot}$, which is close to the typical values reported for dwarf irregulars and ultra diffuse galaxies in the local group \citep{collins14, collins19}. The observational constraints on the stellar mass of the CS progenitor can be obtained from future measurements of $\alpha$ elements abundances. Because [$\alpha$/Fe] is a diagnostic indicator of the intensity of the early star forming activities, which depends on the size of the dwarf progenitor \citep{tolstoy09}. The preferred parameters we obtained from the simulations in this work can shed some light on the nature of the Cetus progenitor dwarf, whereas a thorough numerical research on its properties is required for more quantitative understandings, which needs the kinematic information of more members of this stellar-debris system first.

\begin{table}
\begin{center}
\caption{Initial Conditions of Galaxy Properties}
\bgroup
\def\arraystretch{2}
\scalebox{0.75}{\begin{tabular}{lccccc}
\hline
\hline
          & Milky Way & \multicolumn{4}{c}{Satellite}\\
          &           & D20 & D4 &  d20  & S20    \\
\hline
$M_{\rm vir}$ [M$_{\odot}$]   & $10^{12}$ & 2$\times10^9$ & 4$\times10^8$& 2$\times10^9$ & 2$\times10^9$  \\
$R_{\rm vir}$ [kpc]       & 245     & 20.5        & 11.98 & 20.5    & 20.5         \\
$c$                       & 15.3    & 15           & 15& 15       & 15         \\
$\epsilon_{\rm dm}$ [kpc]  & --      & 0.02        & 0.01 & 0.02    & 0.02         \\
\hline
$M_{\star}$ [M$_{\odot}$] & 7.3$\times10^{10}$  & 2$\times10^8$ & 4$\times10^7$& 2$\times10^8$ & 2$\times10^8$  \\
$f_b$                     & 0.685   & 0     & 0      & 0          & 1        \\
$\epsilon_{\star}$ [kpc]  & --      & 0.01     & 0.005   & 0.01       & 0.01     \\
\hline
$r_{\rm d}$ [kpc]         & 3.0     & 1.24   & 0.25    & 0.69       & --       \\
$h_{\rm d}$ [kpc]         & 0.28    & 0.62   & 0.12   & 0.20       & --        \\ 
$r_{\rm b}$ [kpc]         & 0.2     & --       & --   & --         & 0.6      \\
$r_{1/2}$ [kpc]       & --     & 1.63     & 0.96  & 0.87       &   1.08   \\
$M_{1/2}$ [$M_{\odot}$] & --   & 3.7$\times 10^8$   & 7.4$\times 10^7$  & 2.1$\times10^8$     &   2.6$\times10^8$  \\ 
\hline
\hline
\label{tab:ic}
\end{tabular}}
\egroup
\end{center}

-- Note: In each satellite dwarf , the total number of dark matter particles $N_{\rm dm}$ is 4.5$\times10^{5}$, and that of star particles $N_{\star}$ is 2$\times10^{5}$.
\vspace{0.2cm}%

\end{table}

\begin{table}
\begin{center}
\caption{Initial Conditions of Controlled Runs}
\bgroup
\scalebox{0.75}{\def\arraystretch{2}
\begin{tabular}{lccc}
\hline
\hline
Run &$R_{\rm init}$ ($x$, $y$, $z$)  & $V_{\rm init}$ ($v_{\rm x}$, $v_{\rm y}$, $v_{\rm z}$) &  $T_{\rm inf}$ \\
&kpc&km/s&Gyr\\
\hline

(D20, $\mathcal{O}$1, P8)  & ($-$13.4, $-$35.6, 6.9) & (44.0, $-$36.0, $-$106.0) & 5.05  \\
\hline
(D20, $\mathcal{O}$2, P4)  & (18.8, 25.4, $-$17.0) & ($-$25.6, 83.3, 93.6) & 2.50  \\
(D20, $\mathcal{O}$2, P6)  & ($-$12.8, $-$32.2, 9.4) & (48.3, $-$51.6, $-$107.9) & 3.71   \\
(D20 $\mathcal{O}$2, P8)$\dagger$  & (5.0, 35.5, $-$1.1) & ($-$60.7, 12.1, 114.5) & 4.92   \\
(d20, $\mathcal{O}$2, P8)  & (5.0, 35.5, $-$1.1) & ($-$60.7, 12.1, 114.5) & 4.92   \\
(S20 ,$\mathcal{O}$2, P8)  & (5.0, 35.5, $-$1.1) & ($-$60.7, 12.1, 114.5) & 4.92   \\
(D4, $\mathcal{O}$2, P8)  & (5.0, 35.5, $-$1.1) & ($-$60.7, 12.1, 114.5) & 4.92  \\
\hline
\hline
Sgr & (50.4, 8.5, $-$44.2) & ($-$44.3, 11.0, $-$56.3) & 4.7 \\
LMC &  (33.5, 473.2, 46.2)      &   (2.7, $-$6.4, $-$33.9) & 5.1\\
\hline
\label{tab:orb}
\end{tabular}}
\egroup
\end{center}
\vspace{-0.6cm}%
Note: $\dagger$ denotes the fiducial run.
\vspace{0.2cm}%

\end{table}

\subsection{Interaction with the LMC and the Sgr dwarf}
\label{subsec:int}

In this section, we discuss the effect on the Cetus Stream from the Large Magellanic Cloud (LMC), as well as the possible interaction between the CS progenitor dwarf and the Sgr dwarf.

The LMC is the most massive satellite of the MW, located in the Galactic South. Recent studies show that its presence could have had significant effects on some of the thin stellar streams in our Galaxy. For example, the misalignment between the track and the motion of the Orphan stream detected by \citet{koposov19} can be explained by the perturbation induced by an LMC with mass of $\sim$ 1.4$\times$10$^{11}$M$_{\odot}$ \citep{erkal18}. Such misalignment may also happen with Tucana III \citep{erkal19a}. 

In this work, we set a simple run to evaluate the effect of the LMC on the CS, by combining the potential of the LMC with the MW. The LMC mass is set to be 1.3$\times10^{11}M_{\odot}$, consistent with \citet{penarrubia16, erkal19b}. The LMC initially is situated at the distance of 49.4 kpc, with line-of-sight velocity and proper motion ($v_{\rm los}$, $\mu_{\alpha}^{*}$, $\mu_{\delta}$) = (262.2 km s$^{-1}$, 1.91 mas yr$^{-1}$, 0.229 mas yr$^{-1}$) from \citet{vanderMarel02, kallivayalil13}. Based on these values, the LMC was $\sim$ 60 kpc away from the center of the Cetus progenitor at the most recent apo-center passage of the latter, in which case, the interaction between them should be negligible along the history.

The integrated orbits of the CS progenitor in the combined potential are plotted in the last row of Fig.~\ref{fig:run}. Since the LMC was accreted very recently, the backward orbit of the progenitor is not affected much compared with the previous runs, shown as the orange solid lines in each panel. The change in the forward orbit is substantial under the influence of the LMC (see blue lines). However, the stream would follow the orbit of the progenitor during stripping. Therefore the simulated stream with the LMC is very similar to the fiducial stream as shown in ($b$, $d$) and ($b$, $v_{\rm gsr}$). The different wraps are color coded same as those from the fiducial run shown in Fig.~\ref{fig:o2}. Note that we use a static MW potential, thus the wake of the MW induced by the LMC is not taken into account. This might affect the morphology of the stream by a noticeable amount according to \citet{gomez15,erkal18}, and needs further investigations.

Anther dwarf galaxy which may interact with the CS progenitor is the Sgr dwarf. The footprints of the ``main'' CS and the Sgr stream overlap in the region of the South Galactic Cap at similar distances, as can be seen from \citet{newberg09, koposov12}. However, their orbital poles have different directions \citep{yuan19}, which makes it possible to distinguish them unambiguously. But since the ``main'' CS and the Sgr stream are debris left behind by their progenitors, these two dwarfs may have crossed paths with each other during their history. 

\begin{figure}
\centering
\includegraphics[width=\linewidth]{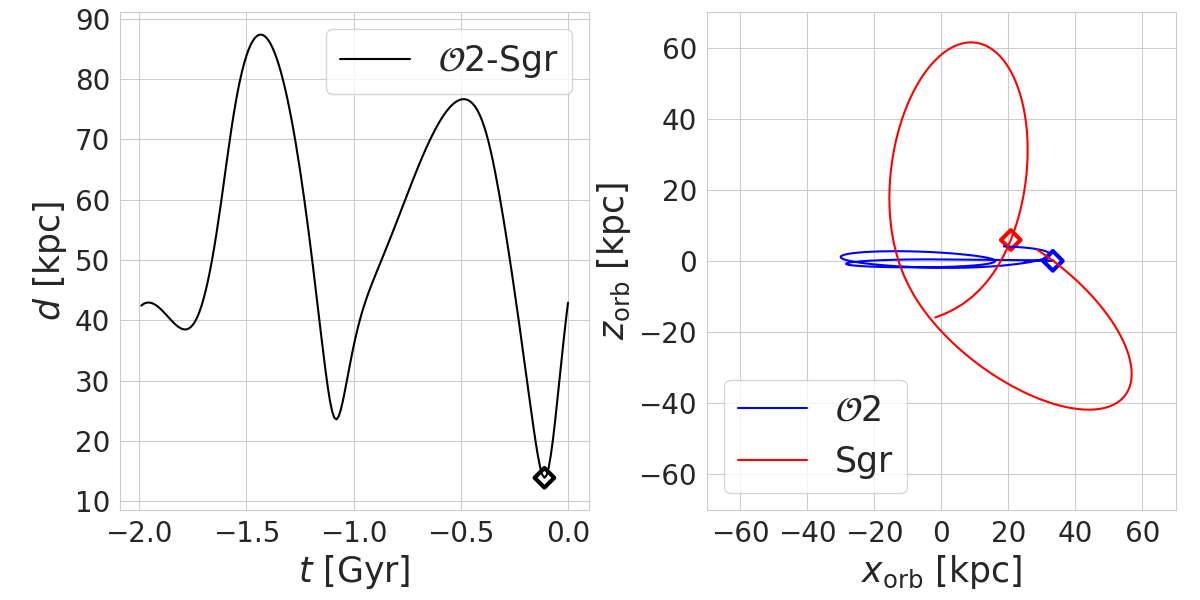}
\caption{The left panel shows the distance between the center of the CS progenitor from the fiducial run and the Sgr dwarf, in the last 2 Gyr. The closest point has a distance of 17 kpc. The right panel shows the Sgr orbit in the coordinate system of $\mathcal{O}$2. The angle between these two orbital poles is about 60$^{\circ}$. }
\label{fig:cssgr}
\end{figure}

We first check the distance between the CS and Sgr dwarf progenitors in the past 2 Gyrs, as shown in the left panel of Fig.~\ref{fig:cssgr}. The closest encounter happened at the distance of 17 kpc, 0.12 Gyr ago. Given the fact that the largest satellite has virial radius $\sim$ 20 kpc, the tidal interactions between these two dwarfs can be safely neglected. The right panel shows the orbit of Sgr in the $\mathcal{O}$2 coordinates system, where the diamond symbols mark the positions of these two progenitors at the closest point. It can be clearly seen that the orbital plane of the Sgr dwarf is almost perpendicular to $\mathcal{O}$2, in which case the tidal stirring between each other is the least significant.

\section{Conclusions}
\label{sec:con}

Motivated by the long time suspicion that NGC 5824 could be the nuclear star cluster of the progenitor dwarf of the Cetus Stream, we perform a series of simulations to investigate this system. Our aim is to recover the general morphology of the stream, which is an extremely challenging task without having identified the progenitor, as is the case for the Sgr dwarf. In order to match all the observational features, for the first time, we unfold the simulated stream along the azimuthal angle $\Psi$, and compare it with different components of the observed stream. Our results show that the dwarf progenitor on the orbit of NGC 5824 is not able to recover the ``main'' CS component from the observations, which is the most densely populated part of the stream. Therefore, Scenario I, that NGC 5824 is the core of this stellar-debris system is disfavored. 

We then try to model the CS system by following the typical orbit from the ``main'' CS members in Scenario II. The resulting stream gives reasonable matches with the observed CS in different phase space, from the fiducial settings with a dark matter halo of 2$\times$10$^9$M$_{\odot}$, a fairly diffuse disc, a long evolution time ($\sim$ 5 Gyr) in the MW of eight apo-center passages. A more thorough investigation of the parameter space and quantitative fittings to the CS stream will be done in the future studies, which requires kinematic and detailed chemical information of more CS members in the first place. Based on the best case scenario explored in the work, the center of the CS progenitor can be traced by the member in its densest part of the stream, and NGC 5824 was situated off-center, which leads to the timing problem for NGC 5824 similar to the Fornax GCs. How to prevent the GC from sinking into the center of the CS progenitor before the system infall to the MW remains unknown, which requires the modeling of the dynamical evolution of the GC in the disrupting dwarf progenitor under the MW potential. Alternatively, NGC 5824 could also be the core of another dwarf that has very similar orbit as the CS progenitor. The verification of this scenario requires the detection of potential stream members in the Southern sky, which will become possible with deep photometric data from the upcoming LSST \citep{lsst09}.

From the current library of streams uncovered by \citet{shipp18} from \citet{des16}, we find that the Palca Stream perfectly matches the simulated stream from the fiducial run, which is very likely the Southern extension of the CS. The kinematic information of the Palca members, especially the line-of-sight velocities are crucial to confirm this hypothesis. Another diffuse structure, Eri-Phe overdensity, situated in the vicinity of the Palca stream \citep{li16}, could also be related to the CS. In the LSST era, we expect to verify these associations, and uncover the full footprint to the CS in the South.

This is the first work dedicated to the modeling of the Cetus Stream and NGC 5824 system, and it shows the power of tailor-made N-body simulations in reproducing stellar streams from dwarf galaxies. This approach could help us unravel the merger tree of our Galaxy, and eventually it could help us set up tailored hydrodynamical simulations of the Milky Way.

\section*{Acknowledgements}

This study is supported by the National Natural Science Foundation of China (NSFC) under grants No. 11988101, 11873052, 11890694, 11973048, 11927804 and National Key R$\&$D Program of China No. 2018YFA0404501, No. 2019YFA0405500. This work is also supported by the Astronomical Big Data Joint Research Center, co-founded by the National Astronomical Observatories, Chinese Academy of Sciences and the Alibaba Cloud. J.C. is supported by the Young Scientists Foundation of JiangSu Province (BK20181110). J.C. and X.K. acknowledge the NSFC fundings (No. 11825303, 11861131006). Z.Y. acknowledges the support from Special Funding for Advanced Users through LAMOST FELLOWSHIP and the Shanghai Sailing Program (Y955051001). I.T.S. acknowledges support from the PIFI Grant n. 2018PM0050. T.S.L is supported by NASA through Hubble Fellowship grant HST-HF2-51439.001 awarded by the Space Telescope Science Institute, which is operated by the Association of Universities for Research in Astronomy, Inc., for NASA, under contract NAS5-26555. This project was developed in part at the 2019 Gaia-LAMOST Sprint workshop, supported by the NSFC under grants 11873034, U1731108 and U1731124.

\bibliography{ms}
\bibliographystyle{apj}

\end{document}